\documentclass[prb,twocolumn,superscriptaddress,showpacs,preprintnumbers,amsmath,amssymb,10pt]{revtex4-1}

\usepackage{color}
\definecolor{red}{rgb}{1,0,0}

\usepackage[T1]{fontenc}
\usepackage[pdftex]{graphicx}
\usepackage{dcolumn}
\usepackage{bm}

\usepackage{mathcomp}

\DeclareGraphicsExtensions{.pdf}   
\usepackage[justification=raggedright,singlelinecheck=false]{caption}

\usepackage{ulem}

\begin{document}

\title{The nature of the electronic band gap in lanthanide oxides}
\author{Roland Gillen}\email{rg403@cam.ac.uk}
\affiliation{Department of Engineering, University of Cambridge, Cambridge CB3 0FA, United Kingdom}
\author{Stewart J. Clark}
\affiliation{Physics Dept, Durham University, Durham, United Kingdom}
\author{John Robertson}
\affiliation{Department of Engineering, University of Cambridge, Cambridge CB3 0FA, United Kingdom}

\date{\today}

\begin{abstract}
Accurate electronic structures of the technologically important lanthanide/rare earth sesquioxides (Ln$_2$O$_3$, with Ln=La,...,Lu) and CeO$_2$ have been calculated using hybrid density functionals HSE03, HSE06 and screened-exchange (sX-LDA). We find that these density functional methods describe the strongly correlated Ln f-electrons as well as the recent G$_0$W$_0$@LDA+U results, generally yielding the correct band gaps and trends across the Ln-period. For HSE, the band gap between O 2p states and lanthanide 5d states is nearly independent of the lanthanide, while the minimum gap varies as filled or empty Ln 4f states come into this gap. sX-LDA predicts the unoccupied 4f levels at higher energies, which leads to a better agreement with experiments for Sm$_2$O$_3$, Eu$_2$O$_3$ and Yb$_2$O$_3$.
\end{abstract}


\maketitle

\section{Introduction}
Lanthanide oxides are a group of compounds whose unique electronic properties allow important applications as catalysts\cite{ganduglia-pirovano-Ovac}, catalyst support, high dielectric constant gate oxides\cite{Afanasev-2004}, and other applications including dopants for lasers, materials for magneto-optic memory and colorants for special glasses. Many properties of lanthanide (Ln) oxides are determined by their semi-core 4f levels. While being mainly localized on the Ln atoms and usually not participating in bonding and electronic conduction, 4f shell electrons are available for optical absorption and can establish strong magnetic order. 

Intriguingly, many physical properties of the lanthanide sesquioxdes Ln$_2$O$_3$ are found to be periodic in the series Ln=La,...,Lu. Available experimental data\cite{borchardt-1963,prokofiev96-gaps} on the band gaps shows a similar periodicity, with four distinct dips being observed for Ce, Eu, Tb and Yb. They attributed the dips for Ce$_2$O$_3$ and Tb$_2$O$_3$ to occupied f-levels entering the forbidden gap above the valence band, and the other two dips to minima in the conduction band energy. In this work, we argue that the dips for Eu$_2$O$_3$ and Yb$_2$O$_3$ arise from empty 4f states entering the gap. On the other hand, the reduction of CeO$_2$ to Ce$_2$O$_3$ by releasing an oxygen atom leads to the transfer of electrons to the 4f orbitals of the two Ce atoms, with the two single occupied 4f levels being pushed down deep into the electronic band gap. The proper description of the 4f electrons at moderate computational cost is thus of greatest importance for the correct prediction of the electronic and magnetic properties of these oxides, allowing for simulations of impurities in systems containing rare-earth elements and catalytic properties.

It is well known\cite{hay06,dasilva07} that the common density functional theory (DFT) approaches of the local-density (LDA) and generalized gradient approximation (GGA) cannot properly describe the 4f electrons in rare-earth compounds. This problem arises from several deficiencies in standard DFT, including the lack of self-interaction cancellation. This leads to an artificial delocalization of the electronic wavefunctions\cite{perdew-sic}, which contributes to the underestimation of electron excitation energies, sometimes closing the band gap completely. Hence, purely local approaches fail to reproduce both the magnitude and the periodicity of the band gaps of lanthanide sesquioxides and predict that Ce$_2$O$_3$ is metallic (LDA) or a small-gap semiconductor (GGA). 

The problem can be partly solved by adding an empirical Hubbard U potential for the 4f orbitals within the LDA+U/GGA+U approach, as in Ref.~\onlinecite{ganduglia-pirovano-2009,loschen2007}, or by explicitly accounting for self-interaction in self-interaction corrected LDA (SIC-LDA)\cite{petit-2005}. 
A more sophisticated treatment is the GW method\cite{Hedin-GW}, where many-body effects are introduced by an (energy-dependent) self-energy term $\Sigma(\vec{r},\vec{r}',\epsilon)\approx iG(\vec{r},\vec{r}',\epsilon)W(\vec{r},\vec{r}',\epsilon)$, with a one-particle Green's function G and a dynamically screened Coulomb interaction W. Inclusion of this self-energy considerably improves the description of excited state properties, but it is expensive if it is carried out in a fully self-consistent way. GW is often used as a  correction to the Kohn-Sham energies using the LDA/GGA wave-functions, either as a pure oneshot correction (GW approximation/G$_0$W$_0$) or by additional self-consistent updating of the Green function (GW$_0$). Jiang \textit{et al.}\cite{jiang2009,jiang-2012} have shown that G$_0$W$_0$ corrections to LDA+U ground states give improved agreement to experiments. However G$_0$W$_0$ is perturbative and cannot be used for total energy and structural optimization, which are valuable for treating defects in catalysis. 

Hybrid exchange-correlation functionals are an alternative, moderate cost method for including non-local electron-electron interactions into a single-particle description. Non-local interactions are included by mixing in a (statically screened) fraction of Hartree-Fock exchange into the LDA/GGA exchange-correlation, effectively cancelling the self-interaction effects. This gives a great improvement in the predicted electronic and magnetic properties\cite{byklein-SX,Seidl-SX,clark-ZnO,clark-sx,pacchioni-2008,hse03,hse06,eyert-2011,guo-ti203,gillen-magiso}. Crucially, they are generalized Kohn-Sham functionals\cite{Seidl-SX} that allow a self-consistent calculation of ground state properties giving a well-defined total energy leading to atomic forces and geometries, in contrast to G($_0$)W$_0$.
We shall see that the different incorporation of non-local exchange into standard local functionals leads to interesting variations in the predicted band structures of the 15 Ln$_2$O$_3$ oxides and of CeO$_2$. 

In this paper, we present accurate electronic properties of Ln-oxides, greatly improving over LDA and GGA for all investigated materials and yielding results on par with the GW@LDA+U values\cite{jiang2009,jiang-2012} but at a fraction of the computational cost.

\section{Method}
For this paper, we used two hybrid functionals; HSE and sX-LDA, as implemented in the CASTEP computational package\cite{castep}. HSE, which exists in two different variants, HSE03\cite{hse03} and HSE06\cite{hse06}, substitutes 25\% of error-function screened PBE exchange by the same amount of Hartree-Fock exchange. The two differ in the screening parameter $\omega$ used for the range separation, specifically $\omega$=0.158\,bohr$^{-1}$ for HSE03 and $\omega$=0.106\,bohr$^{-1}$ for HSE06.

On the other hand, sX-LDA\cite{Seidl-SX, byklein-SX,clark-sx} divides LDA into short-range and long-range part and replaces the short-range exchange part by 100\% Hartree-Fock exchange, screened by the Thomas-Fermi function. In principle, the Thomas-Fermi wave vector $k_{\mbox{\tiny TF}}$, which depends on the average charge density, is used as the inverse screening length $k_s$. An alternative approach is to use a fixed screening length by setting k$_s$ to the screening length minimizing the total energy per electron in a homogeneous electron gas. In contrast to the bare Thomas-Fermi screening, this approach takes, too a certain extent, electron-electron interactions into account, which leads to a general increase of the screening length
. We thus used a fixed value of $k_s$=0.76\,bohr$^{-1}$, which works well for most s-p semiconductors.

We modeled all 15 lanthanide sesquioxides by the hexagonal (A-type) structure of space group P-3m, while CeO$_2$ was modeled by the three atom primitive cell of the cubic flourite structure (space group Fm-3m). The atomic cores were described by norm-conserving pseudopotentials generated by the OPIUM code. We treated the (4f,5s,5p,5d,6s) states as valence states by plane waves with a cutoff energy of 750 eV. Reciprocal space integration was performed by k-point grids of 4x4x3 and 4x4x4 k-points in the Brillouin zone of Ln$_2$O$_3$ and CeO$_2$, respectively. We optimized the geometry at the GGA level preserving the symmetry of the system until the pressure on the unit cell was below 0.01\,GPa and the residual forces between the atoms were below 0.01\,eV/Å. The densities of states was calculated by a grid of 10x10x10 equally spaced points in the Brillouin zone.

A final word about the magnetic order and convergence in our calculations. As the electronic spin density is allowed to change during the electronic relaxation, the local minima in the energy surface correspond to different magnetic states of the material. We initialised the spin density to a given magnetic state, an antiferromagnetic order in case of the lanthanide sesquiocides, in an attempt to find a particular energy minimum. For exchange-correlation functionals of the local density, this technique is usually successful in obtaining the particular local minimum of interest. However, in the case of hybrid functionals a significant part of the Hamiltonian is constructed from the wave function, which is not included in such a magnetic state initialisation. Numerically, this can cause instability in the electronic solver or allows the system to relax into a different magnetic state than the intended one. This problem was noted previously for the spin-polarized LDA+U method\cite{korotin-1996,ylvisaker-2009}, which is an on-site approximation to the Hartree-Fock exchange.
\section{Results and Discussion}
\subsection{CeO$_2$}
\begin{figure}[t!]
\centering
\begin{minipage}{0.48\textwidth}
\includegraphics*[width=0.95\textwidth]{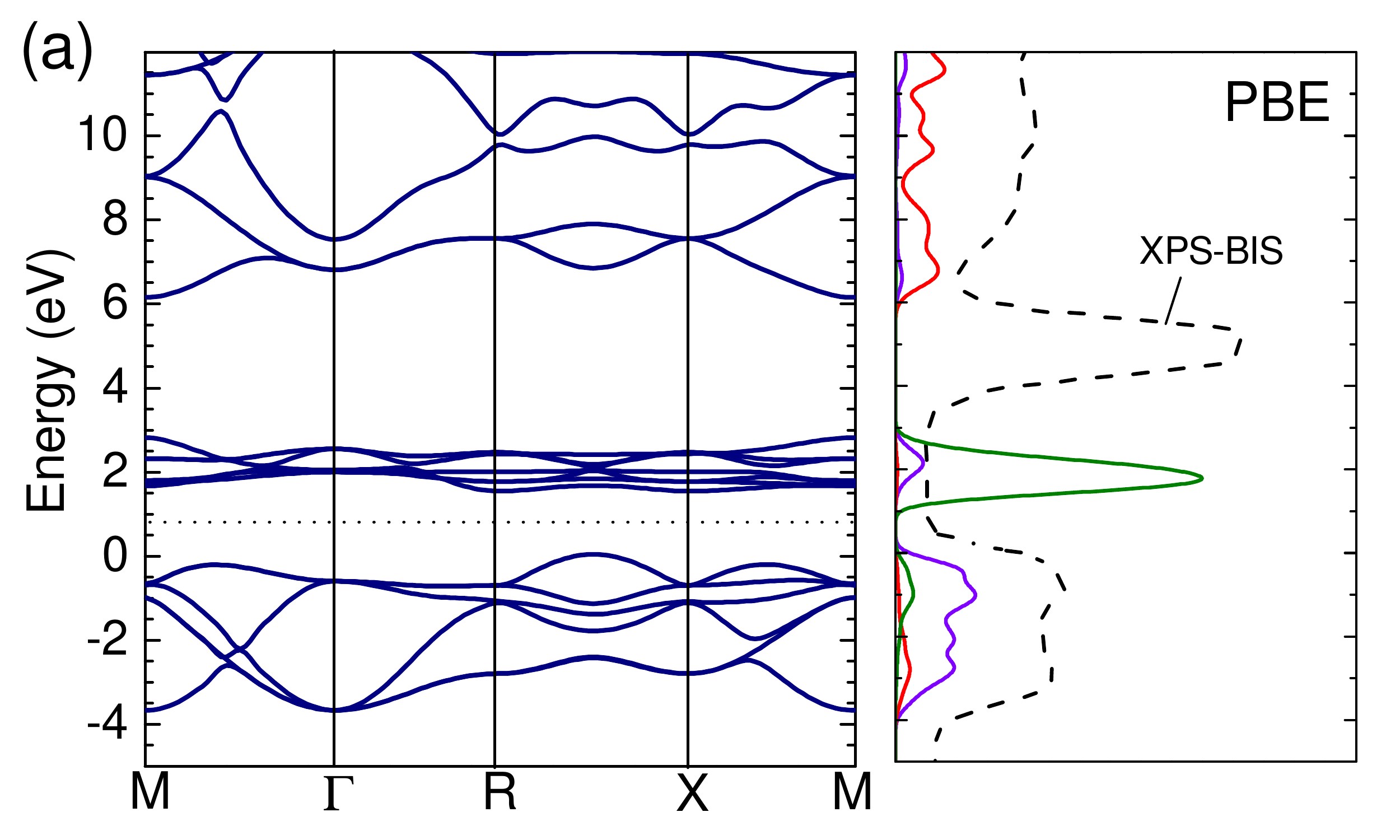}
\end{minipage}
\\
\begin{minipage}{0.48\textwidth}
\includegraphics*[width=0.95\textwidth]{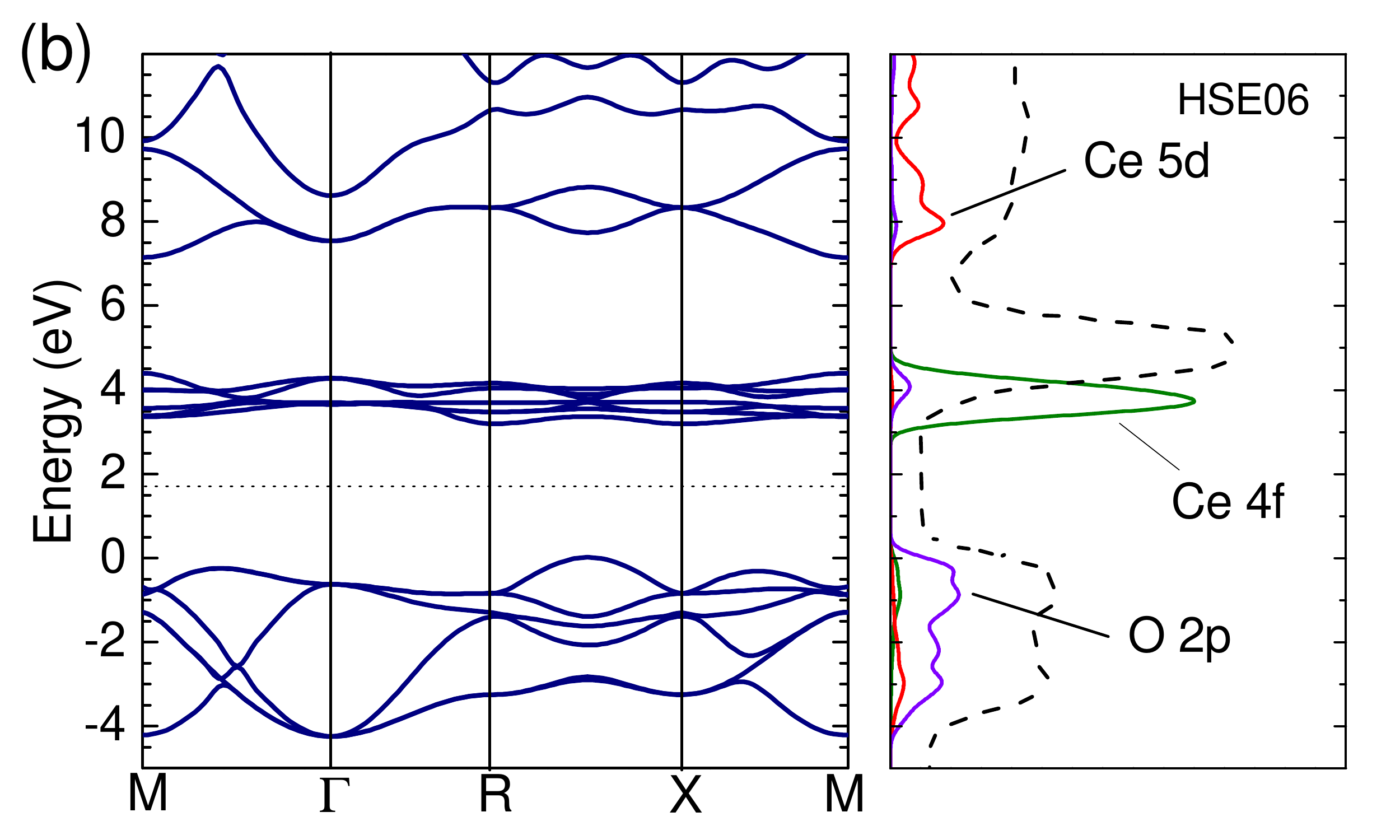}
\end{minipage}
\\
\begin{minipage}{0.48\textwidth}
\includegraphics*[width=0.95\textwidth]{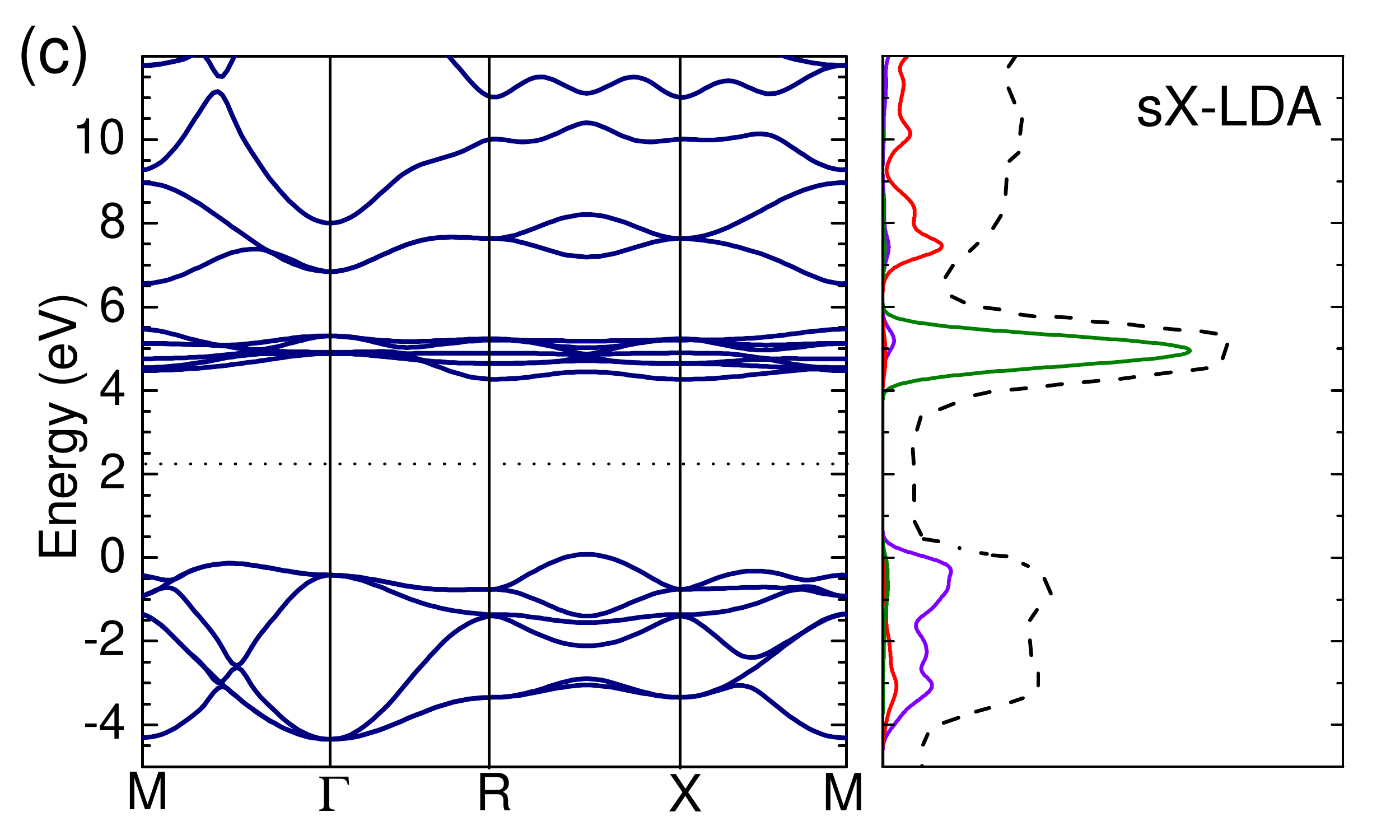}
\end{minipage}
\caption{\label{fig:CeO2} (Color online) Band structures and (partial) density of states (DOS) of CeO$_2$ as calculated from the (a) PBE, (b) HSE06 and (c) sX-LDA exchange-correlation functionals. The dashed (grey) lines in the DOS plots depicts the experimentally measured DOS from Ref.~[\onlinecite{wuilloud-ceo2}] for comparison. We chose the valence band maximum as the zero energy. The dotted line in the band structure plots depicts the Fermi energy.
}
\end{figure}
As a first step, we evaluate the electronic structure of CeO$_2$. Ce donates all its 6s and 5d valence electrons to the oxygen atoms, while the ground state of the 4f orbitals was established to be a mixed valence state of 4f$^0$ and 4f$^1$\underline{v} (\underline{v} is a ligand hole) configurations\cite{mixedvalence}. This makes the valence band top O 2p in character, with a contribution from 4f states\cite{wuilloud-ceo2}. The lowest conduction band consists mainly of empty Ce 5d states starting at $\approx$6\,eV above the valence band edge. Bremsstrahlung isochromat spectroscopy (BIS)\cite{wuilloud-ceo2} and x-ray absorption spectroscopy\cite{mullins1998} show a sharp peak below the conduction band, which was attributed to empty Ce 4f levels taking part in electron addition processes. Experimentally, the size of the band gap of CeO$_2$ is debated, but believed to be in the range of 5.5-8\,eV\cite{marabelli1987,mullins1998,wuilloud-ceo2,pfau1994}.
\begin{table}
\begin{ruledtabular}
\caption{\label{tab:gaps-VBM} Minimum band gaps (in eV) of CeO$_2$ as obtained from PBE, 
HSE, sX-LDA, B3LYP, PBE0 and G$_0$W$_0$@LDA+U calculations and experiments. 
}
\begin{tabular*}{\columnwidth}{@{\extracolsep{\fill}} c  c c c }
Method$\slash$Property&p-f band gap&p-d band gap&VBM\\
\hline
PBE&1.5&6.2&3.7\\
HSE03&3.1, 3.5\cite{hay06,dasilva07}&6.9, 7\cite{hay06,dasilva07}&4.3\\
HSE06&3.4&7.2&4.3\\
sX-LDA&4.2&6.5&4.4\\
B3LYP\cite{kullgren2010}&3.3&7.65&-\\
PBE0\cite{kullgren2010}&3.94&8.08&-\\
G$_0$W$_0$@LDA+U\cite{jiang2009}&4&6&4.4\\
Exp&3-3.5\,eV\cite{wuilloud-ceo2,mullins1998}&5.5-8\cite{marabelli1987,mullins1998,wuilloud-ceo2,pfau1994}&$\approx$\,4.5-5\cite{wuilloud-ceo2,mullins1998}\\
\end{tabular*}
\end{ruledtabular}
\end{table}
Figure~\ref{fig:CeO2} compares the calculated band structures and partial density of states (DOS) of CeO$_2$ for the local exchange-correlation functional PBE and the two hybrid functionals HSE06 and sX-LDA. 
The PBE valence band DOS agrees fairly well with experiment, although the predicted valence band width of 3.7\,eV is slightly less than the experimental value of $\approx$4.4\,eV. On the other hand, the failure of PBE becomes very clear for the empty states. The 5d conduction band edge of 6.2\,eV lies at the lower end of the experimental band gaps and, most strikingly, PBE vastly underestimates the energy of the 4f levels compared to the photoemission peak. Splitting the DOS into orbital contributions clearly shows the effect of the self-interaction, a considerable transfer of charge from the O$^{2-}$ ions into the 4f orbitals due to mixing of O 2p and Ce 4f states at the valence band top, and the weakly dispersive gap states. This is reflected in the Mulliken populations of -0.63\,e for the oxygen atoms and 1.27\,e for the cerium atom.

Replacing the short-range part of the PBE exchange by Hartree-Fock exchange largely improves on the PBE results. For HSE06, the valence band is stretched by 0.5\,eV to 4.3\,eV, and is now in good agreement with the experimental DOS from x-ray photoelectron spectroscopy (XPS)\cite{wuilloud-ceo2}. The empty Ce 4f states experience a strong shift by $\approx$2 eV, but still remain below the experimental peak. The higher energy of the Ce 4f states renders mixing of Ce 4f and O 2p states unfavourable, and corresponds to a slight delocalization of the empty 4f states and an accumulation of charge at the oxygen atoms, clearly shown in the partial DOS and the Mulliken population (O: -0.76\,e, Ce: 1.53\,e). The non-local exchange further produces a rigid shift of the conduction bands, opening the O 2p-Ce 5d band gap to 7.2 eV, well within the range of experimental band gaps. Compared to the studies of Hay \textit{et al.}\cite{hay06} and da Silva \textit{et al.}\cite{dasilva07}, our HSE03 calculations predict the Ce 5d states at slightly higher energies and the Ce 4f states at slightly lower energies, respectively. Indeed, it shows that the prediction of the 4f peak (and to a lesser extend for the 5d levels) is quite sensitive to the reference state of the Cerium pseudopotential. We thus specifically generated and used a tetravalent cerium potential for our calculations on CeO$_2$. We believe this causes the differences between our study and previous values.

sX-LDA pushes the 4f-levels to higher energies than HSE06. This leads to a wide O 2p-Ce 4f gap of 4.2 eV, which is in perfect agreement with the experimental position of the dominant peak and with G$_0$W$_0$@LDA+U data\cite{jiang2009}. We further find a p-d band gap of ~6.5\,eV and a wider valence band of 4.4\,eV. The weaker prediction of itinerant states, in this case the 5d levels, but more rigorous treatment of localized states compared to HSE06 has been observed for a variety of materials, \textit{e.g.} transparent conducting oxides\cite{gillen-TCO}, and appears to be caused by the differences in short- and middle-range exchange between the two exchange-correlation functional types. 
Table~\ref{tab:gaps-VBM} summarizes the resulting band gaps and valence band widths.

\subsection{Ln$_2$O$_3$}
\begin{table}
\begin{ruledtabular}
\caption{\label{tab:gaps-Ln2O3} Comparison of minimum band gaps (in eV) of the series of lanthanide sesquioxides as obtained from various theoretical methods and experiments..
}
\begin{tabular*}{\columnwidth}{@{\extracolsep{\fill}} c c c c c c c }
Material&HSE03&HSE06&sX-LDA&G$_0$W$_0$\cite{jiang-2012}&GW$_0$\cite{jiang-2012}&Exp\\
\hline
La$_2$O$_3$&4.5&5.1&5.5&4.95&5.24&5.5\footnotemark[1], 5.34\footnotemark[2]\\
Ce$_2$O$_3$&2.37&3.38&1.75&1.5&1.29&2.4\footnotemark[1]\\
Pr$_2$O$_3$&3.5&3.77&3.8&2.86&2.82&3.9\footnotemark[1], 3.5\footnotemark[2]\\
Nd$_2$O$_3$&4.32&4.63&4.65&4.5&4.7&4.7\footnotemark[1], 4.8\footnotemark[2]\\
Pm$_2$O$_3$&4.5&4.8&5.6&5.25&5.41&\\
Sm$_2$O$_3$&3.2&3.4&4.8&4.38&5.22&5\footnotemark[1]\\
Eu$_2$O$_3$&2.25&2.5&4&2.77&3.48&4.4\footnotemark[1]\\
Gd$_2$O$_3$&4.95&5.26&4.85&4.89&5.3&5.4\footnotemark[1]\\
Tb$_2$O$_3$&3.7&4&&3.81&3.74&3.8\footnotemark[1]\\
Dy$_2$O$_3$&4.7&4.9&&4.41&4.24&4.9\footnotemark[1]\\
Ho$_2$O$_3$&4.83&5.08&&4.68&5.12&5.13\footnotemark[1]\\
Er$_2$O$_3$&4.93&5.3&&4.78&5.22&5.3\footnotemark[1], 5.49\footnotemark[2]\\
Tm$_2$O$_3$&4.55&4.8&&4.73&5.15&5.4\footnotemark[1], 5.48\footnotemark[2]\\
Yb$_2$O$_3$&2.8&3.24&4.5&3.23&4.7&4.9\footnotemark[1],  5.05\footnotemark[2]\\
Lu$_2$O$_3$&4.86&5.136&4.85&4.66&4.99&5.5\footnotemark[1],  5.79\footnotemark[2]\\
\end{tabular*}
\footnotetext[1]{Ref.~\onlinecite{prokofiev96-gaps}}
\footnotetext[2]{Ref.~\onlinecite{kimura-2000}}
\end{ruledtabular}
\end{table}
\begin{figure}[t!]
\centering
\begin{minipage}{0.9\columnwidth}
\includegraphics*[width=0.95\textwidth]{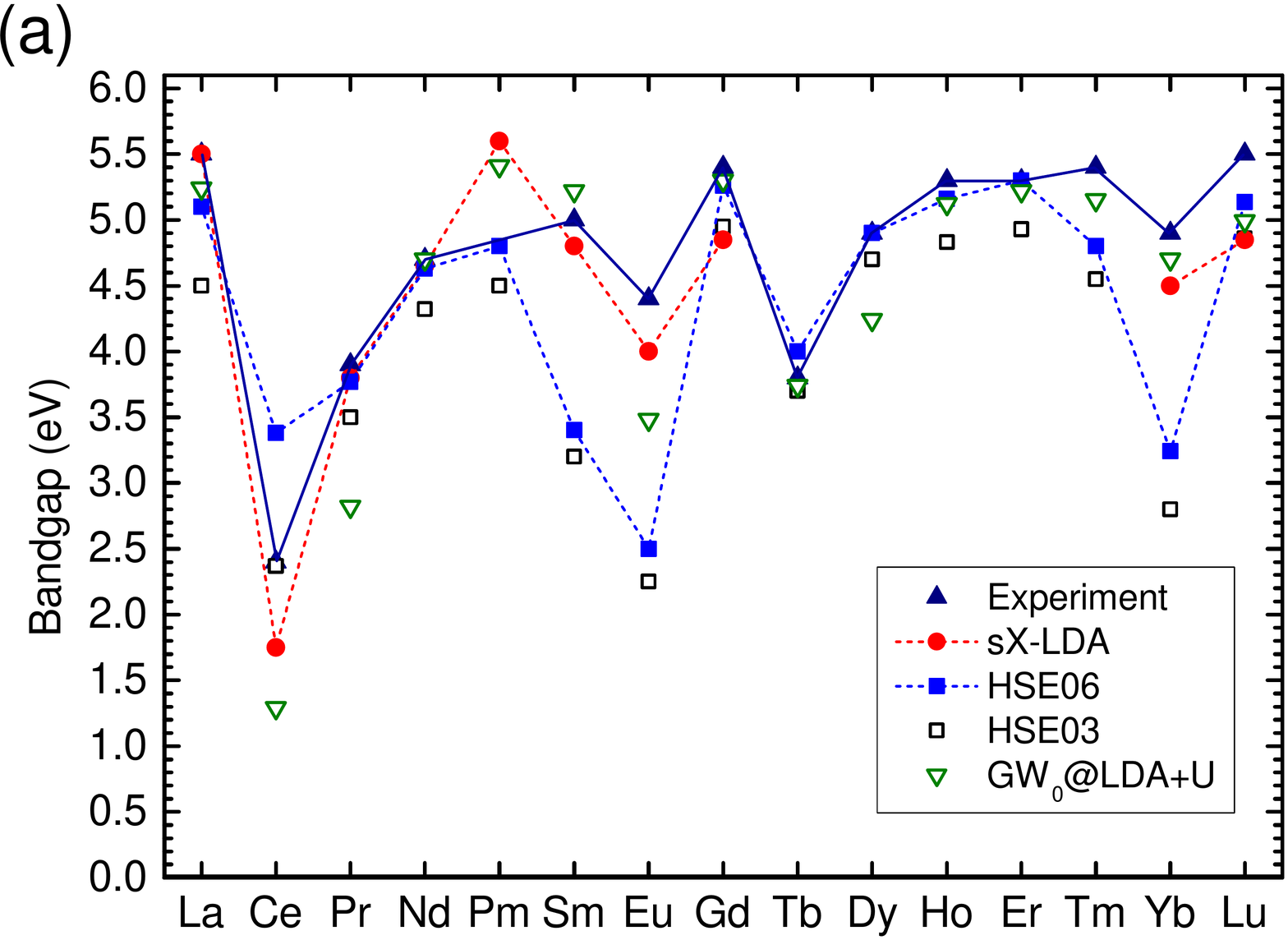}
\end{minipage}
\\
\begin{minipage}{0.9\columnwidth}
\includegraphics*[width=0.95\textwidth]{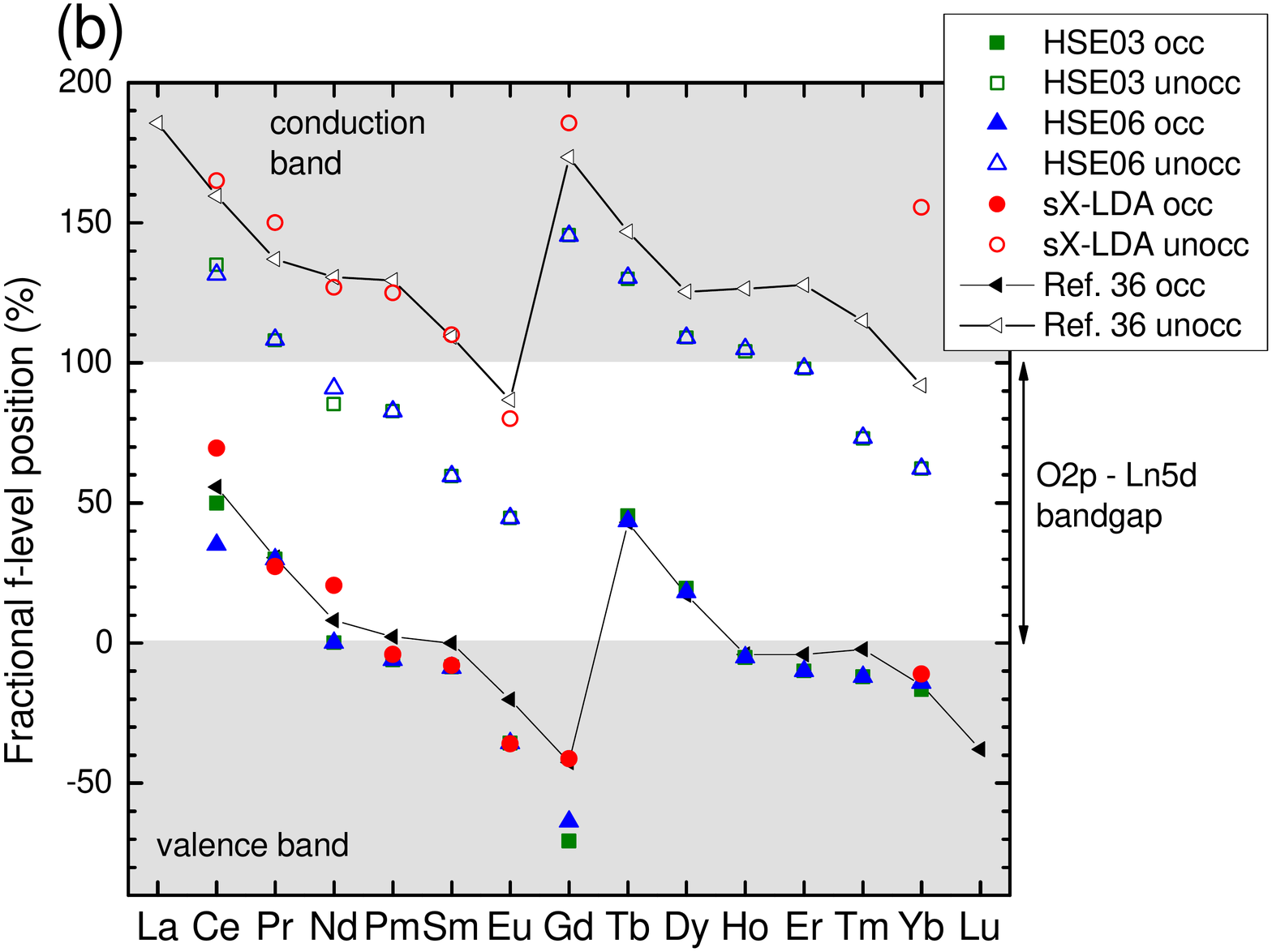}
\end{minipage}
\caption{\label{fig:Ln2O3-levels} (Color online) (a) Comparison of minimum electronic band gaps in the series of lanthanide sesquioxides Ln$_2$O$_3$ as obtained from calculations using the HSE03 (empty black squares),HSE06 (filled blue squares) and sX-LDA (filled red circles) calculations with experimental data (filled upward triangles) and the GW$_0$@LDA+U results from Ref.~[\onlinecite{jiang-2012}] (empty green downward triangles). (b) Corresponding positions of occupied (filled symbols) and unoccupied (empty symbols) f levels relative to VBM and CBM from HSE and sX-LDA calculations. The empty and filled black triangles connected by lines are the 4f levels derived from experiments in Ref.~\onlinecite{vanKolk-2006}.}
\end{figure}
\begin{figure*}[t!]
\centering
\begin{minipage}{\textwidth}
\centering
\includegraphics[width=0.12\textwidth]{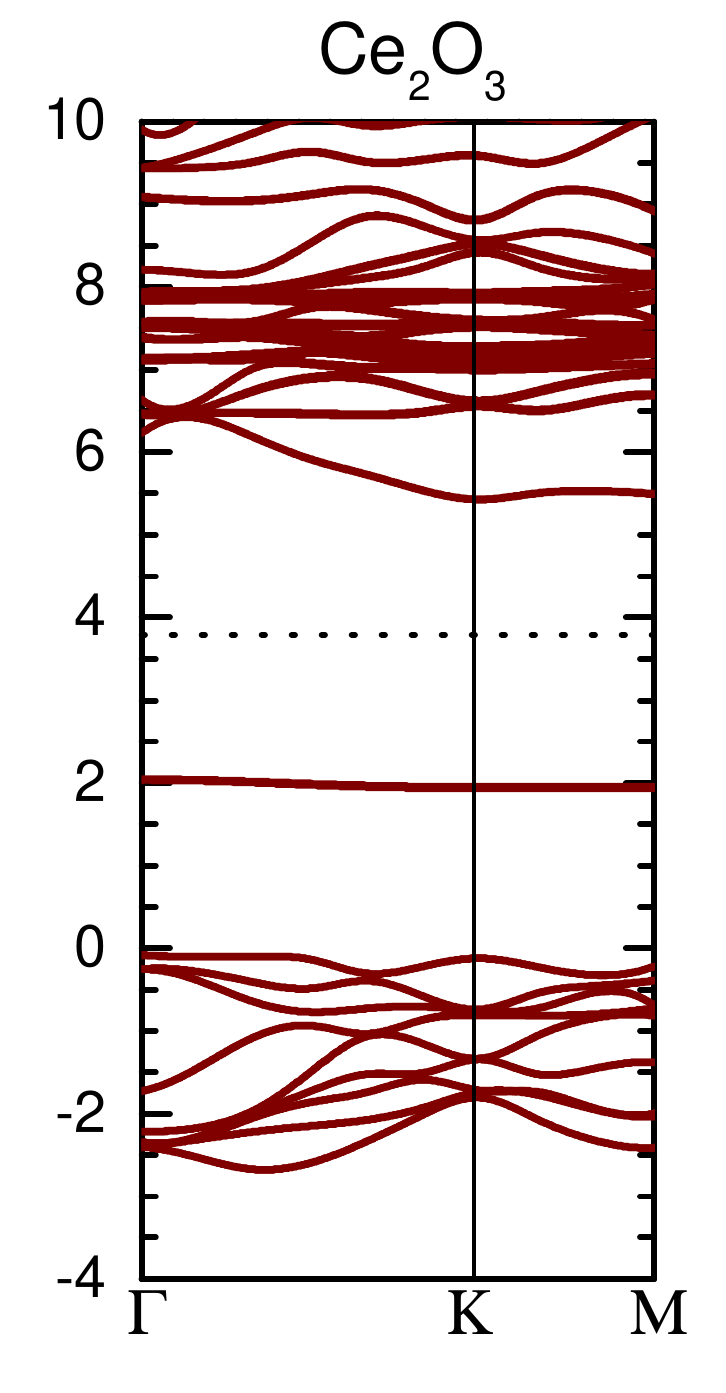}
\includegraphics[width=0.12\textwidth]{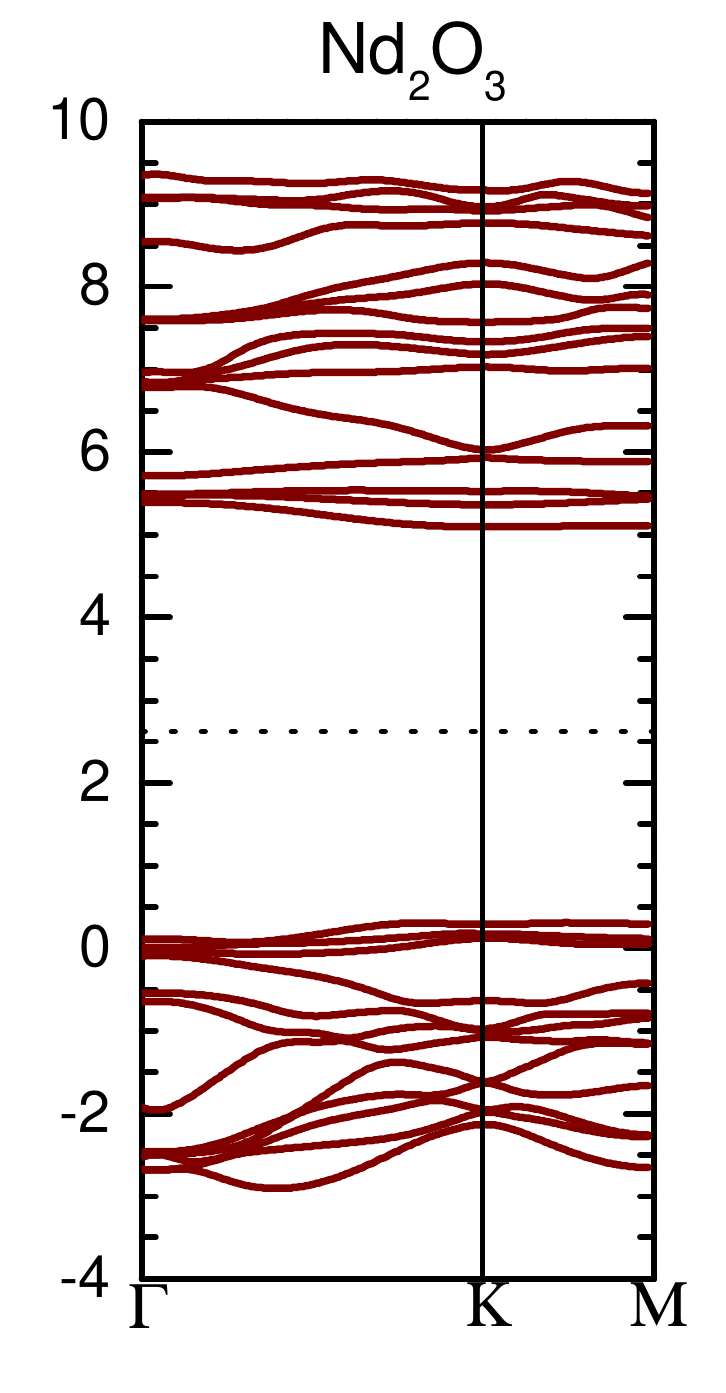}
\includegraphics[width=0.12\textwidth]{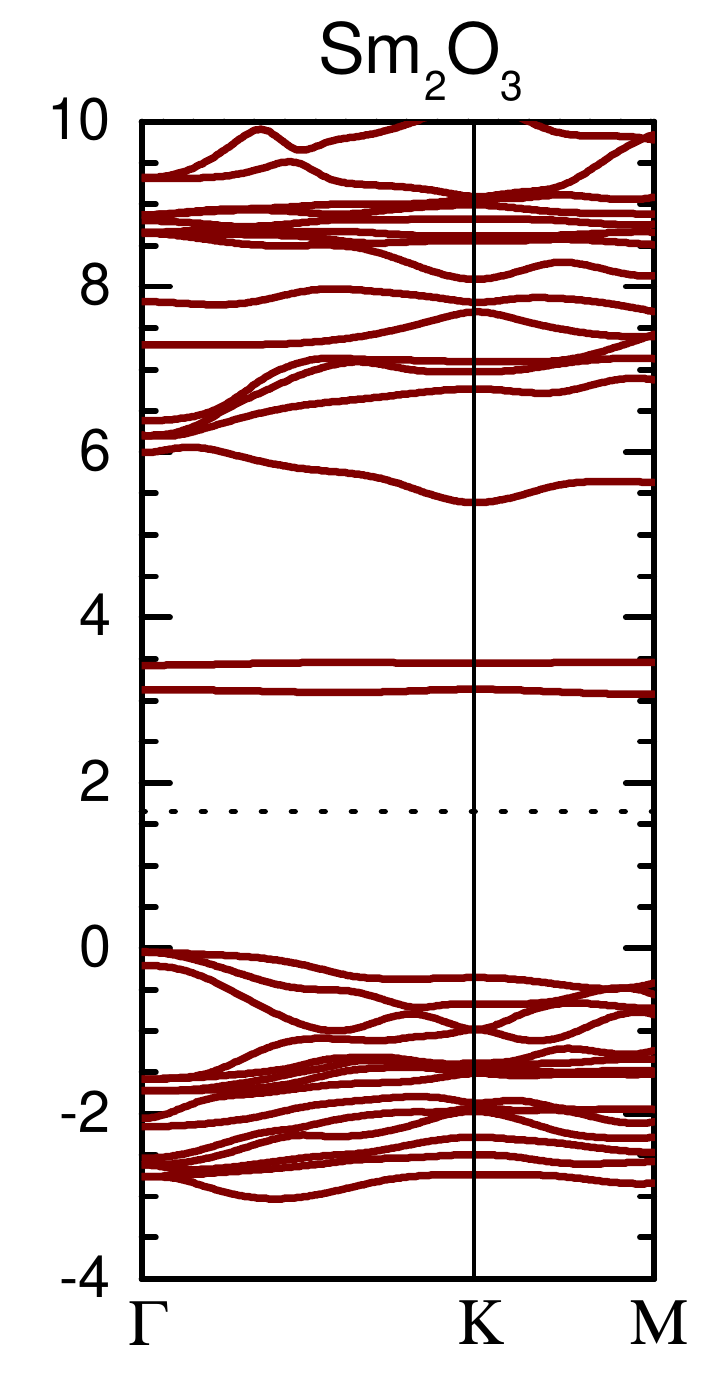}
\includegraphics[width=0.12\textwidth]{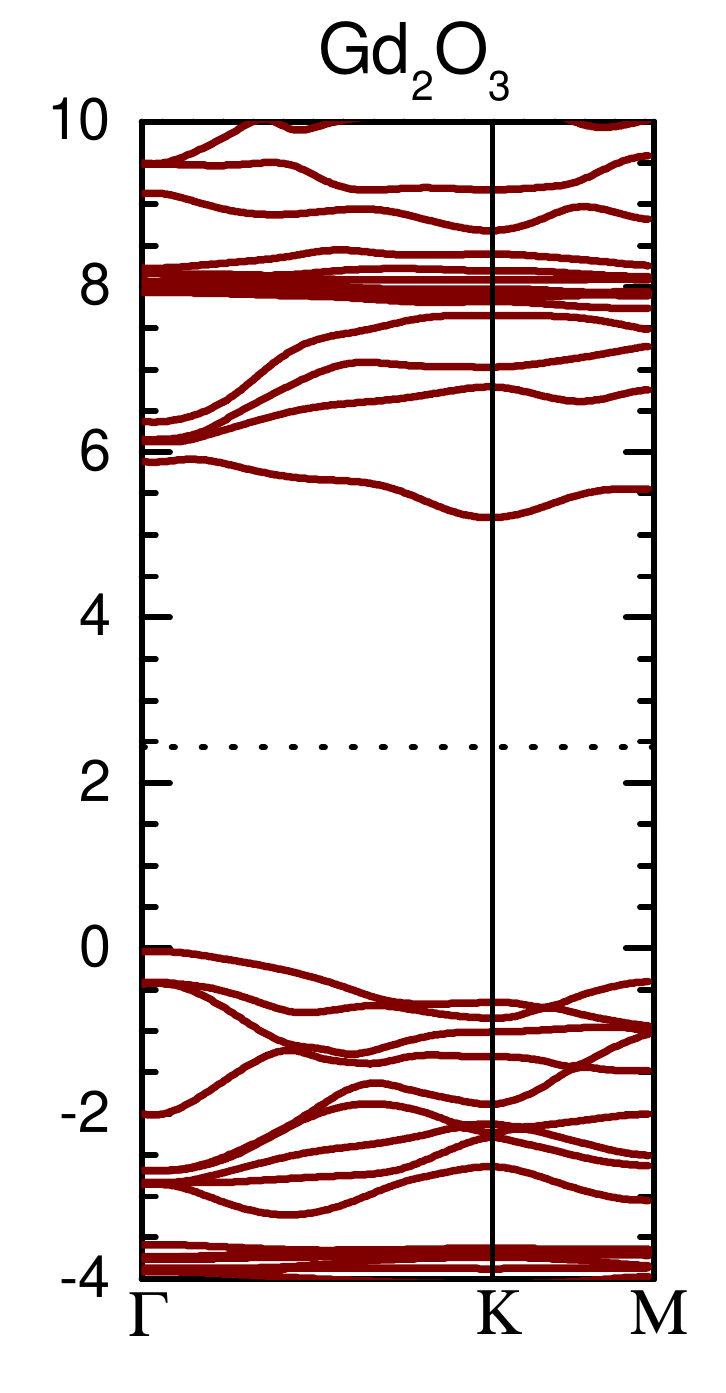}
\includegraphics[width=0.12\textwidth]{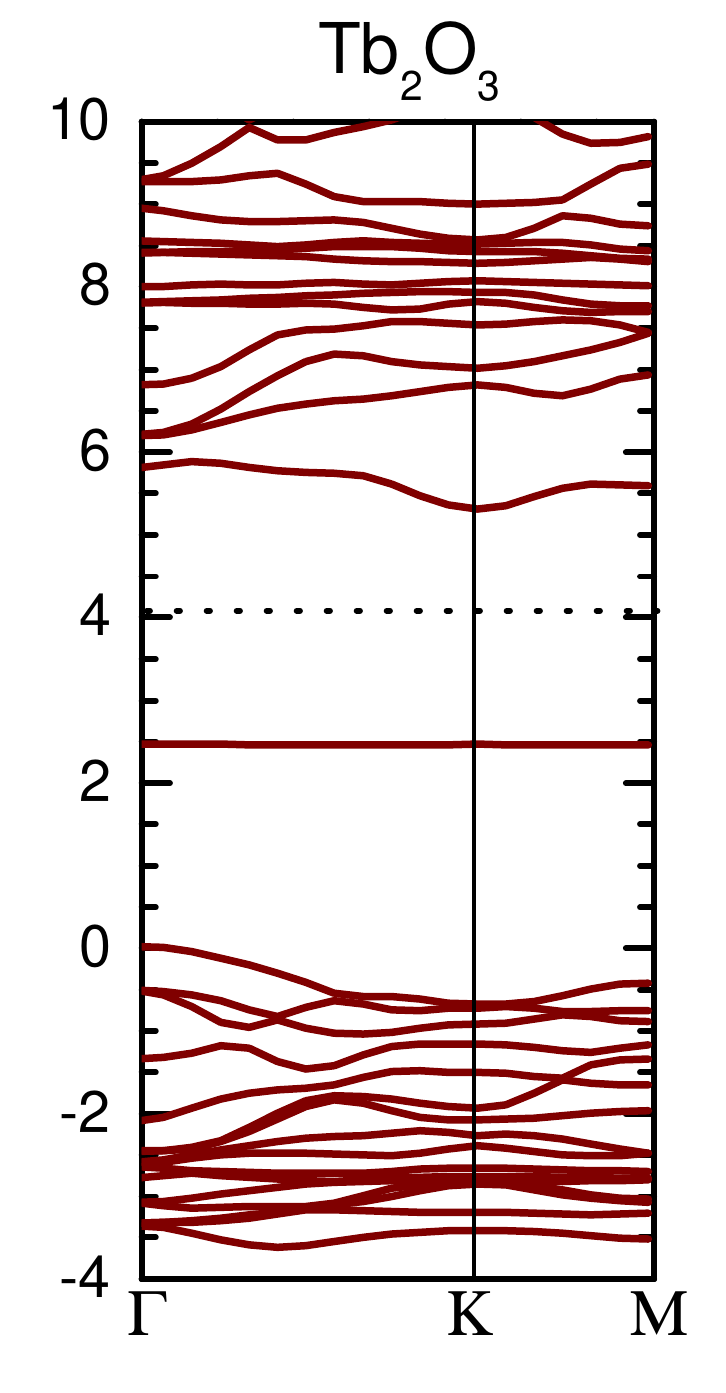}
\includegraphics[width=0.12\textwidth]{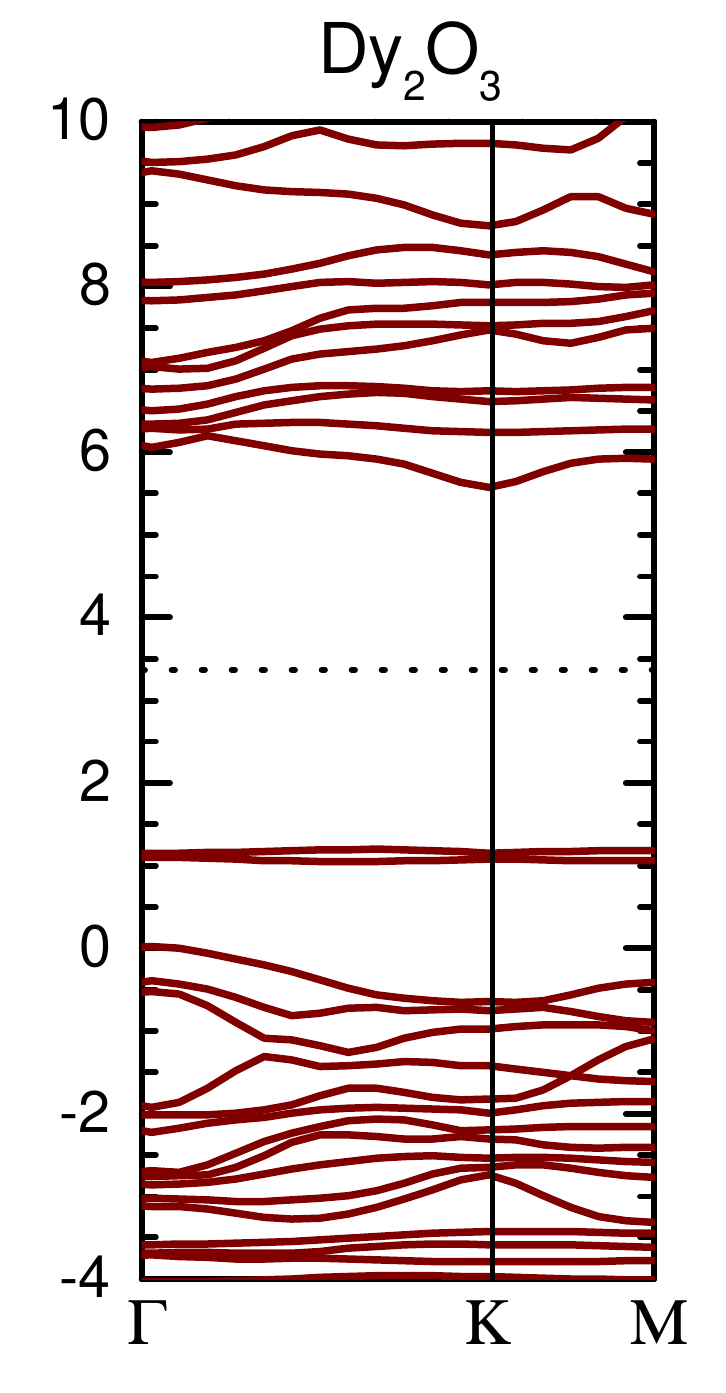}
\includegraphics[width=0.12\textwidth]{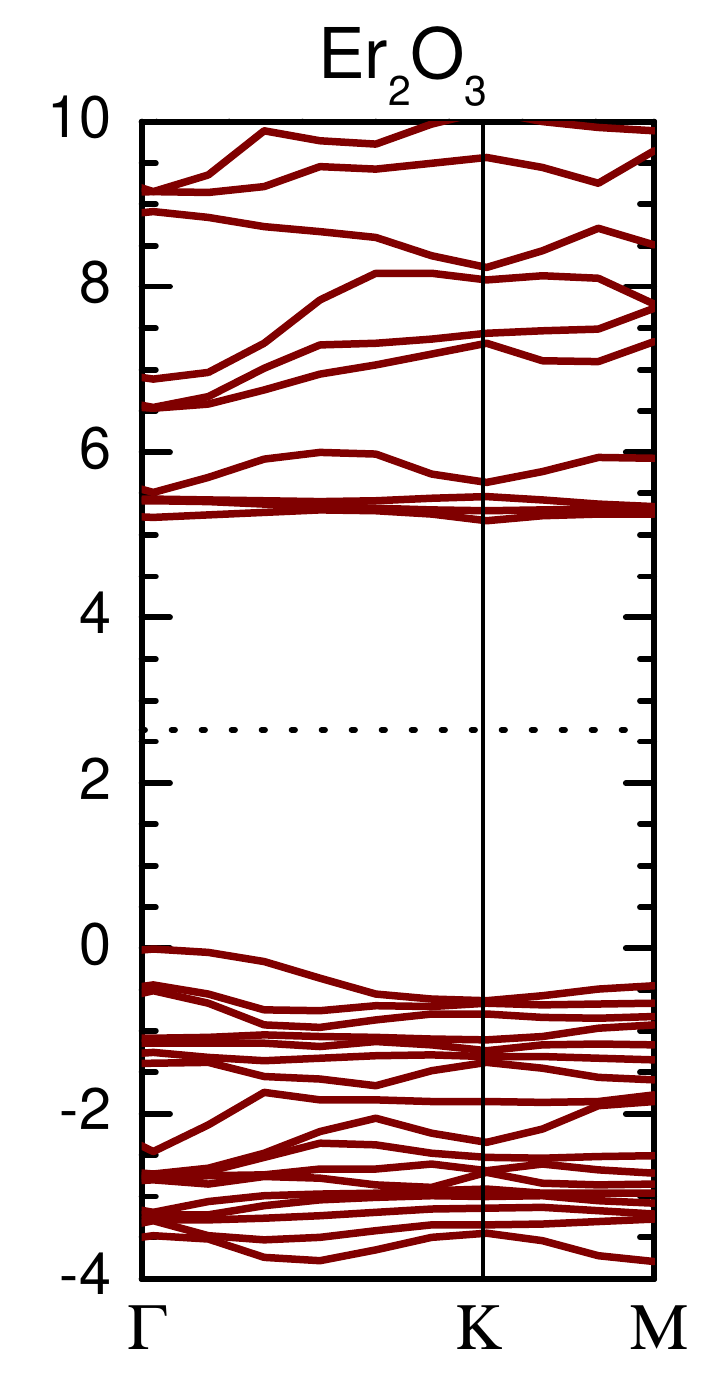}
\includegraphics[width=0.12\textwidth]{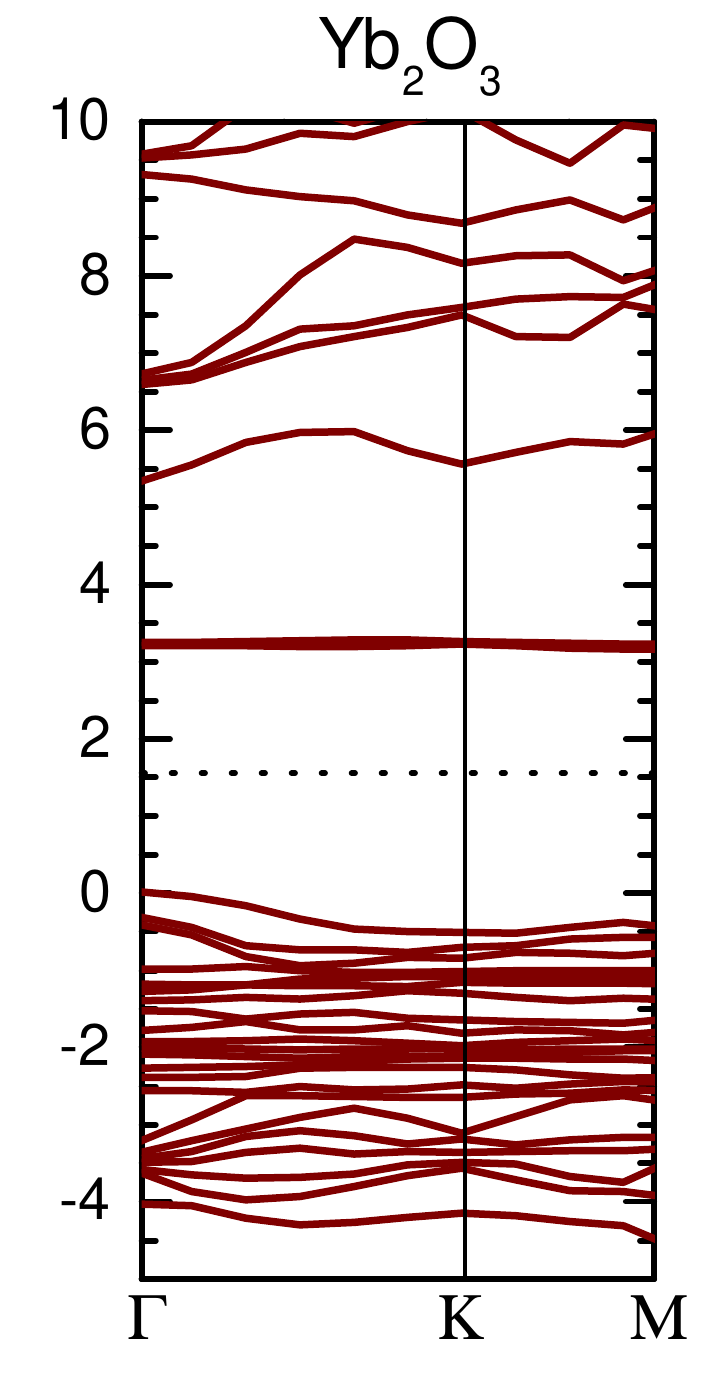}
\end{minipage}
\\
\begin{minipage}{\textwidth}
\includegraphics[width=0.12\textwidth]{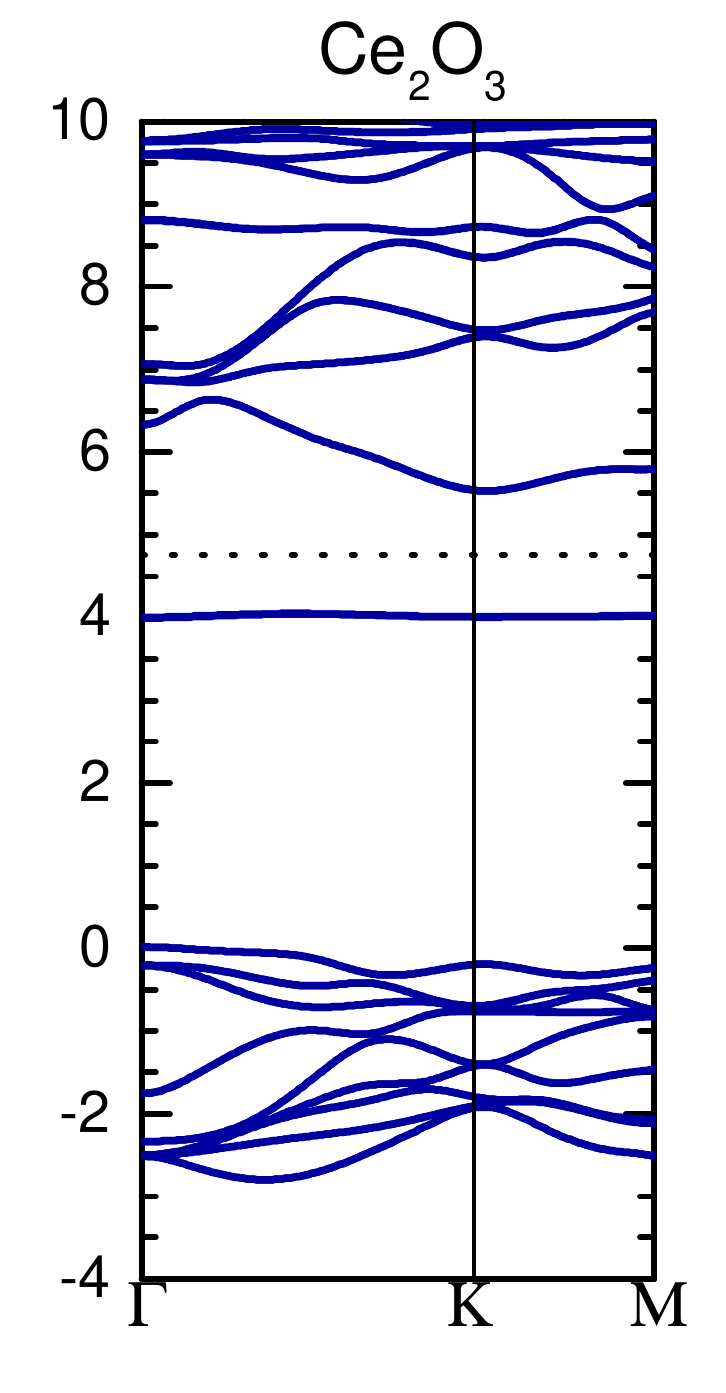}
\includegraphics[width=0.12\textwidth]{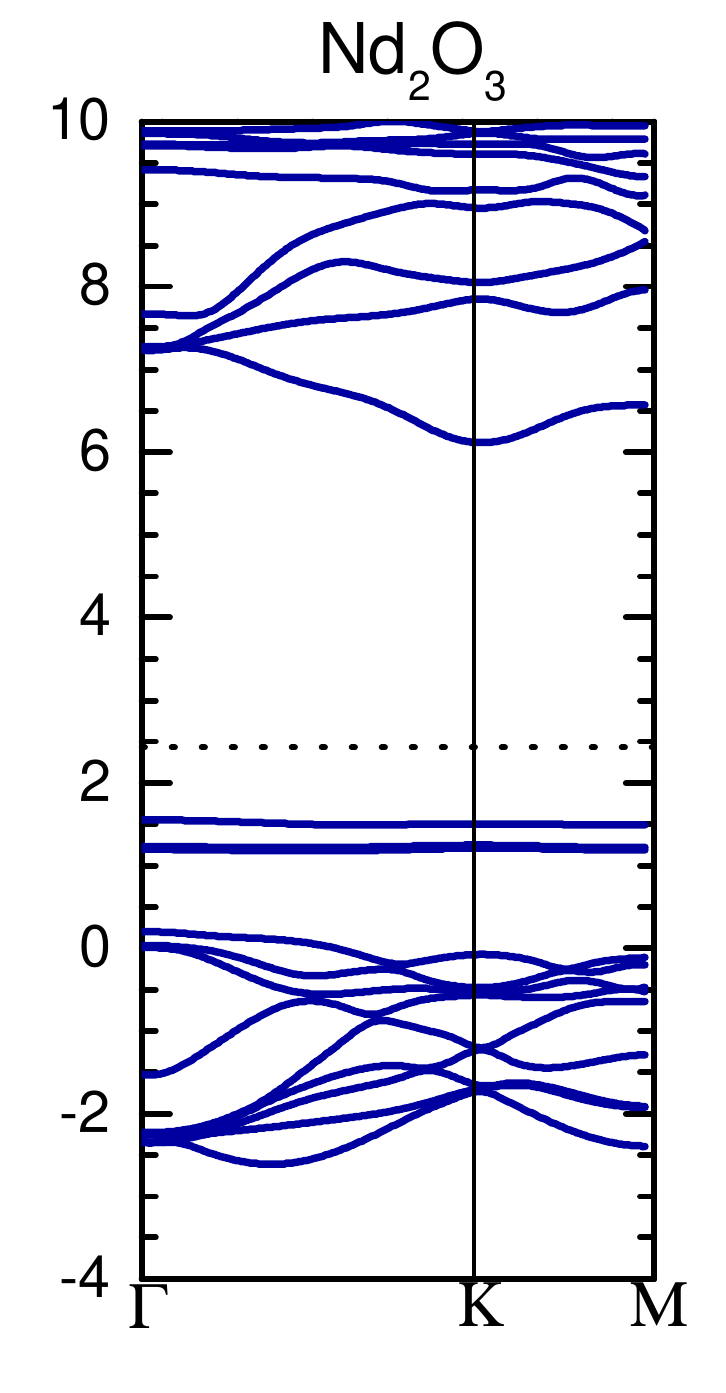}
\includegraphics[width=0.12\textwidth]{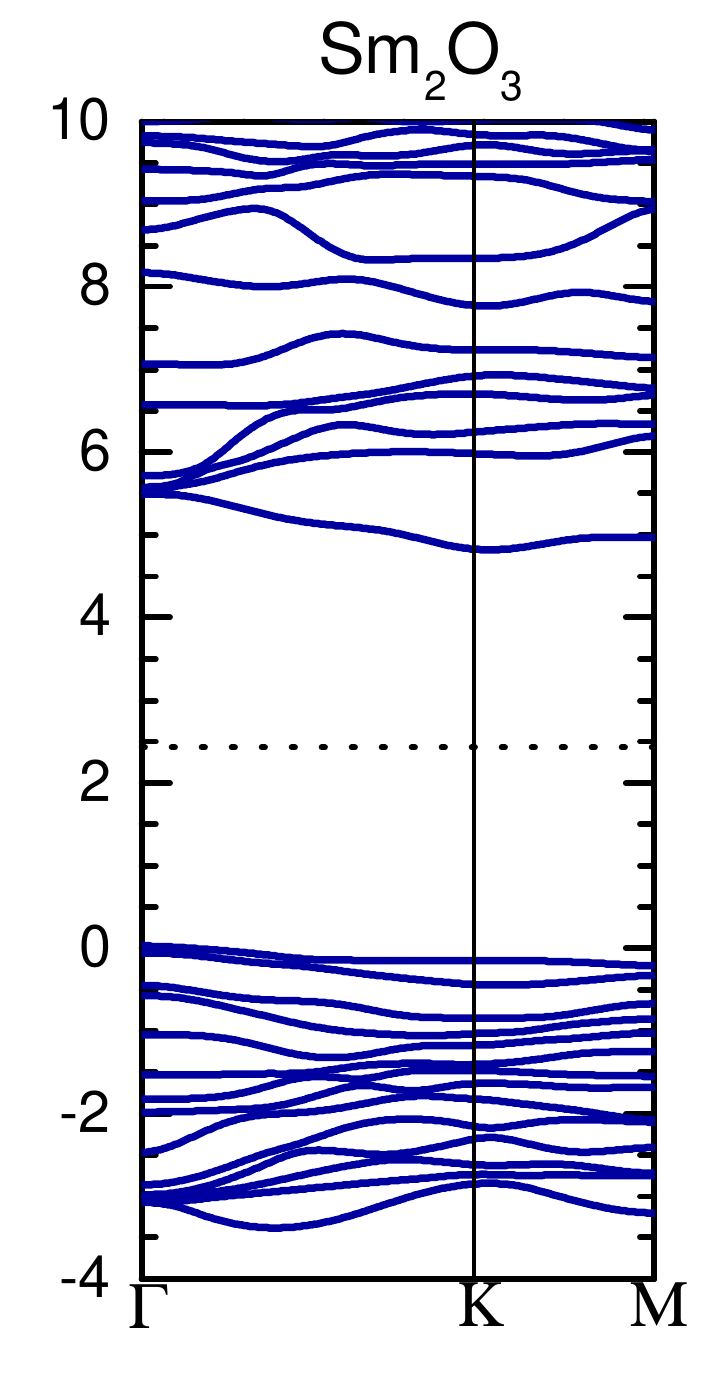}
\includegraphics[width=0.12\textwidth]{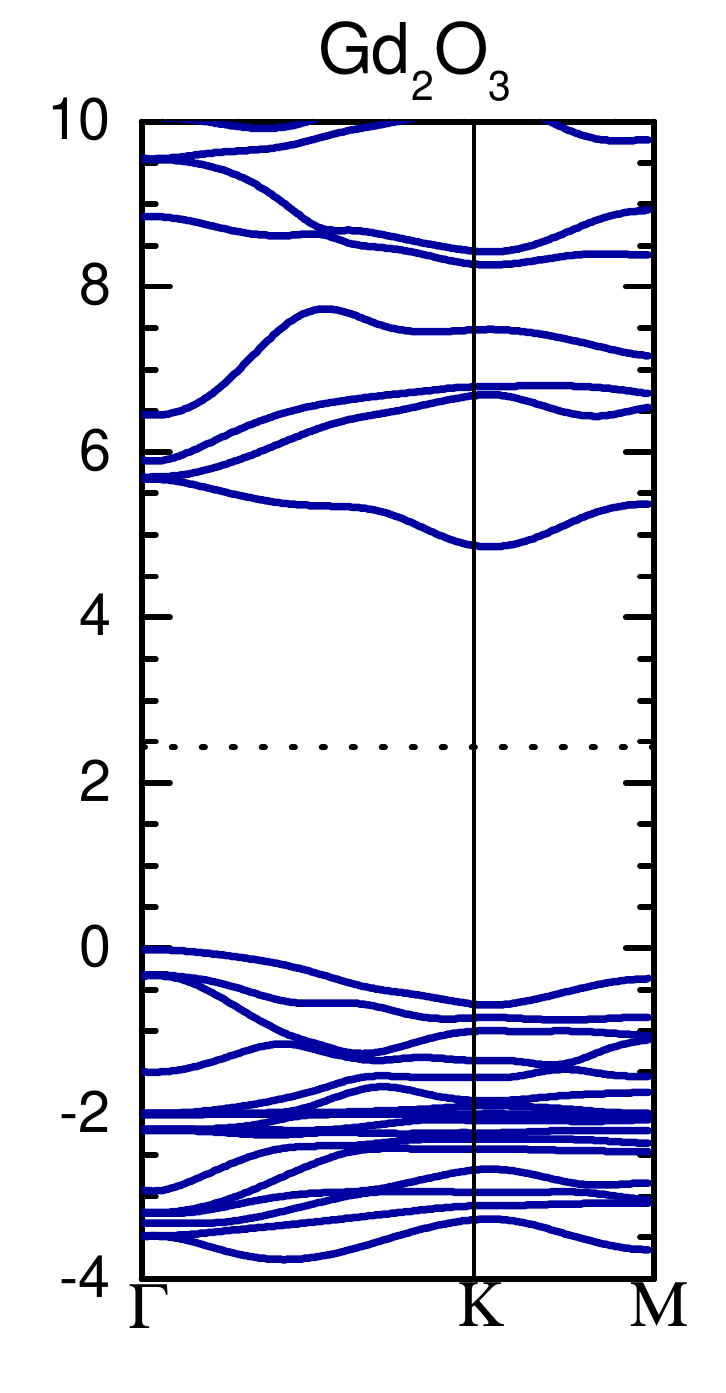}
\includegraphics[width=0.12\textwidth]{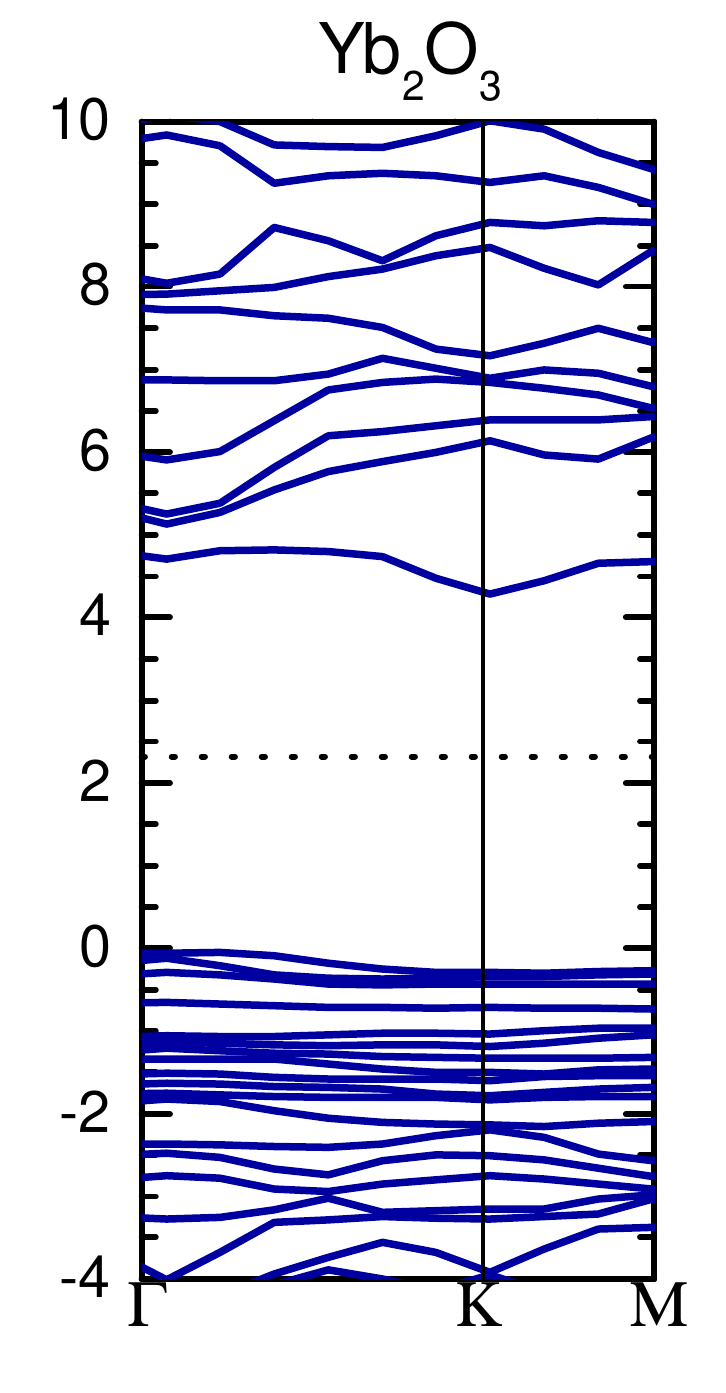}
\end{minipage}
\caption{\label{fig:Ln2O3-bands} (Color online) Band structures of selected lanthanide oxides Ln$_2$O$_3$ from (a) HSE06 and (b) sX-LDA calculations, respectively. We chose the maximum of the oxygen 2p bands as the zero energy for a better comparison of the various band structures. As in Fig.~\ref{fig:CeO2}, the dotted lines depict the Fermi energy.}
\end{figure*}
Similar predictions for the f-electrons are also found for lanthanide sesquioxides Ln$_2$O$_3$ with Ln=La,...,Lu. Experimental measurements of the optical band gaps show an unusual periodicity across the series with distinct dips for Ce, Eu, Tb and Yb oxides\cite{borchardt-1963,prokofiev96-gaps}. While LDA+U calculations predict the dips in the band gaps, they are not capable of reproducing the detailed trends and band gap sizes. Jiang \textit{et al.}\cite{jiang-2012} recently reported that G$_0$W$_0$ and GW$_0$ corrections to LDA+U ground states improve the predictions and are in close agreement with experiments. 

Fig.~\ref{fig:Ln2O3-levels}~(a) shows the minimum band gaps derived from our band calculations using the three hybrid functionals HSE03, HSE06 and sX-LDA compared to the GW$_0$ and experimental values. Clearly, the statically screened short-range Hartree-Fock exchange in the hybrid functionals has a similar effect as the GW corrections and gives significant improvements on the local functionals and LDA+U, both for experimental trends and band gap sizes. We find that the two HSE functionals restore the periodicity of the band gaps with the four observed dips and yield results in good agreement with the experimental values for almost all investigated lanthanide oxides, with sX-LDA performing even slightly better in the first half of the series. Unfortunately, we were not able to obtain sX-LDA band gaps for most oxides in the second half of the series.

Fig.~\ref{fig:Ln2O3-levels}~(b) shows the predicted energy of the highest filled and empty f-levels relative to the valence band maximum and conduction band minimum for the three hybrid functionals. The energies of the f-level energies decrease monotonically across the series from Ce to Gd oxide, from within the conduction band to deep into the valence band. The f-shells reach half-filling at Gd$_2$O$_3$ and the process then repeats. The minimum band gaps correspond to one of four different kinds of transitions, depending on the lanthanide, and in good qualitative agreement with conductivity measurements\cite{lal-1988}. 

For the first elements in the series, the fundamental band gap arises from intra-atomic transitions from the occupied Ln 4f states to the Ln 5d states, which form the bottom of the conduction band. At some point of the series, Nd for HSE03/HSE06 and Eu in case of sX-LDA, empty 4f states begin to to fall into the main O 2p-Ln 5d band gap, while the filled 4f states are pushed into the valence band. The minimum band gap then changes to an inter-atomic transition from O 2p states to empty Ln 4f states. We find an exception for Nd$_2$O$_3$, where HSE predicts the highest occupied 4f states slightly above the O 2p bands, leading to an intra-atomic 4f$\rightarrow$4f transition. However, we note that this transition is symmetry-forbidden in photoexcitation processes. For the following oxides in the series, Pm$_2$O$_3$ and Sm$_2$O$_3$, all three hybrid functionals predict a considerable contribution from 4f electrons to the valence band edge. The same behaviour is found in the second half of the series for Ho$_2$O$_3$-Tm$_2$O$_3$.
Our observations for these materials are in good agreement with a recent study of van Kolk \textit{et al.}\cite{vanKolk-2006}, who constructed an energy level diagram similar to Fig.~\ref{fig:Ln2O3-levels} by fitting an empirical model to experimental data. For La, Lu and Gd (and Pm, Sm and Yb in case of sX-LDA), there are no f-bands in the gap and the fundamental transition is O 2p$\rightarrow$Ln 5d.

The HSE functionals systematically predict the f-bands to lie at lower energies in the forbidden gap than GW\cite{jiang2009,jiang-2012} corrections on LDA+U and sX-LDA (see also the band structures in Fig.~\ref{fig:Ln2O3-bands}). For the oxides with 4f$\rightarrow$5d transitions, this automatically leads to the better HSE06 band gaps, provided that the sizes of the forbidden gap from HSE06 and G($_0$)W$_0$ are comparable. On the other hand, the lower energies of empty f-levels in HSE will underestimate the minimum band gaps of oxides with predicted O 2p $\rightarrow$ Ln 4f transitions, particularly Sm$_2$O$_3$, Eu$_2$O$_3$, Tm$_2$O$_3$ and Yb$_2$O$_3$. This leads to considerable discrepancies with the experimentally measured values for these oxides.
\begin{figure}[t!]
\centering
\includegraphics*[width=0.95\columnwidth]{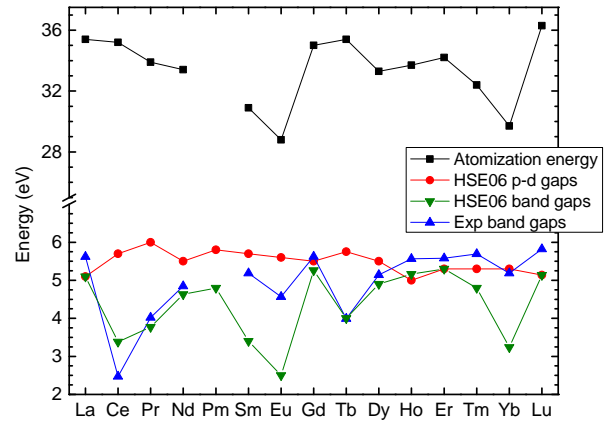}
\caption{\label{fig:atomization} (Color online) Comparison of the trends of the experimentally measured atomization energy (black squares) and band gaps (blue triangles) of 14 lanthanide sesquioxides, taken from Ref.~[\onlinecite{prokofiev96-gaps}] with the values of the band gaps between oxygen 2p and lanthanide 5d states (red circles) and minimum band gaps (green triangles) as predicted from HSE06 calculations.}
\end{figure}
Prokofiev \textit{et al.}\cite{prokofiev96-gaps} attributed the lower optical band gaps of Eu$_2$O$_3$ and Yb$_2$O$_3$ to inter-atomic O 2p$\rightarrow$Ln 5d/4f transitions, being consistent with the observed minima in the atomization energy for these oxides. Their reasoning was that inter-atomic optical excitations, such as O 2p to Ln 5d or O 2p to Ln 4f transitions, were equivalent to a partial breaking of the Ln-O bonds and should thus correlate with the experimental bond/atomization energies. The observed dips in the band gap trends could thus also correspond to minima in the forbidden gap between the O 2p valence band and the Ln 5d conduction band. To resolve this matter for our calculations, we plotted the O 2p$\rightarrow$Ln 5d band gaps as obtained from HSE06 over the whole series and compared them with the atomization energies in Fig.~\ref{fig:atomization}. 
We find that our HSE06 O 2p$\rightarrow$Ln 5d band gaps are fairly independent of the lanthanide and do not exhibit pronounced minima for Eu$_2$O$_3$ and Yb$_2$O$_3$. The observed band gap minima in the HSE functionals hence arise purely from empty f-states entering the gap, without contribution from a lowered 5d conduction band.

Compared to HSE, sX-LDA shows a different qualitative behavior for some of the oxides. It is clear from Fig.~\ref{fig:Ln2O3-bands}~(b) that, as in CeO$_2$, sX-LDA generally places the empty f-levels at significantly higher energies than HSE (and G$_0$W$_0$@LDA+U); they are always within the bulk conduction band for the first half of the lanthanide series, except for Eu$_2$O$_3$. Thus the minimum band gaps in sX-LDA for Pm$_2$O$_3$, Sm$_2$O$_3$ and Yb$_2$O$_3$ are O 2p$\rightarrow$Ln 5d in nature and have values closer to experiment than HSE and G$_0$W$_0$@LDA+U. 

We note that the stronger energetic splitting of occupied and unoccupied semi-core states compared to HSE03 and HSE06 is a known behaviour of sX-LDA\cite{gillen-TCO,gillen-magiso}. While both functional types are conceptually similar in the sense that they both screen the included non-local exchange component in a range-separation scheme, sX-LDA incorporates 100\% of Thomas-Fermi screened Hartree-Fock exchange, whereas HSE includes only 25\% Hartree-Fock exchange, but with a weaker Error function screening and a larger screening length compared to sX-LDA.
The very strong contribution of short-range Hartree-Fock exchange within sX-LDA particularly affects more localized orbitals, such as the $4f$ states in the present lanthanide oxides, and leads to a stronger shift of the corresponding energy levels. We believe that the better description of 4f states within sX-LDA can be traced back to the differences in short-range exchange, while the combination of weaker screening and a smaller portion of Hartree-Fock exchange in HSE is more suited for itinerant electrons, such as in 'classic' s-p semiconductors.

Self-consistently updating the G in the GW corrections, on the other hand, has a similar effect as the larger short-range contribution in sX-LDA compared to HSE. Particularly the unoccupied 4f levels experience a significant energetic up-shift into the conduction band compared to the 'one-shot' G$_0$W$_0$ correction, leading to greatly improved band gaps Sm$_2$O$_3$, Eu$_2$O$_3$ and Yb$_2$O$_3$, at the cost of additional computational effort. The quality of the sX-LDA predictions in general is further underlined by the excellent agreement between our sX-LDA calculations, the GW$_0$@LDA+U levels and empirical model from Ref.~\cite{vanKolk-2006}, particularly for the empty 4f levels.

The effect of energetically higher empty 4f levels is somewhat counteracted by the energetically lower onset of the Ln 5d bands for most oxides along the series, similar to the case of CeO$_2$. While this keeps most of the band gaps on HSE level, it is not clear how well sX-LDA would perform for Ho$_2$O$_3$, Er$_2$O$_3$ and Tm$_2$O$_3$ in the second half of the series, where the band gaps should be O 2p$\rightarrow$Ln 5d in nature. 

Judging from the observations from the oxides of the first half of the series and CeO$_2$, the predictions of sX-LDA for the second half of the series might be considerably lower than the corresponding HSE values. It might thus be interesting to test the performance of a combination of increased screening length and decreased Hartree-Fock ratio to obtain an intermediate solution of the O 2p $\rightarrow$ X 5d gap from HSE and the superior prediction of the empty 4f shells from sX-LDA.

\section{Conclusion}
In summary, the performance of the three hybrid functionals HSE03, HSE06 and sX-LDA for the prediction of the electronic properties of lanthanide oxides is found to be generally comparable to the computationally expensive GW$_0$@LDA+U and G$_0$W$_0$@LDA+U approaches, in particular the nature of the band gap. For lanthanide sesquioxides, we vastly improve on the incorrect ground states of LDA+U calculations, and obtain both the correct trends and the correct sizes of the band gaps for most of the studied oxides. The hybrid functionals used also give the correct ionic charges and correct hybridization of localized f states with the extended O 2p states, which are essential for a good description of catalytic properties.



%

\end{document}